\documentclass[final]{cvpr}

\usepackage{times}
\usepackage{epsfig}
\usepackage{graphicx}
\usepackage{amsmath}
\usepackage{amssymb}
\usepackage{subfigure}
\usepackage{overpic}

\usepackage{enumitem}
\setenumerate[1]{itemsep=0pt,partopsep=0pt,parsep=\parskip,topsep=5pt}
\setitemize[1]{itemsep=0pt,partopsep=0pt,parsep=\parskip,topsep=5pt}
\setdescription{itemsep=0pt,partopsep=0pt,parsep=\parskip,topsep=5pt}

\usepackage[pagebackref=true,breaklinks=true,colorlinks,bookmarks=false]{hyperref}

\def\cvprPaperID{159} 
\def\confYear{CVPR 2020}

\graphicspath{{figures/}}

\begin{document}


\title{Extension: Adaptive Monte Carlo Sampling with Implicit Radiance Field}

\author{Yuchi Huo \\
}

\maketitle
\section{Adaptive Sampling and Reconstruction}\label{sec:Introduction}

This paper aims to explore and summarize the state-of-the-art progress in Monte Carlo adaptive light field sampling and reconstruction using deep reinforcement learning. When using the Monte Carlo algorithm for global illumination drawing, if the sample points generated by path tracing are insufficient, the drawing result will contain a lot of noise, which will seriously affect the usability of the result. In order to solve this problem, there are two common solutions: one is to denoise and filters the drawing results through neural networks or other methods, which can be regarded as a post-processing reconstruction scheme\cite{huo2021survey}; this article introduces the second method is to guide the process of generating Monte Carlo samples from path tracing in the sampling process, so as to improve the quality of the final drawing results, which is a scheme for optimizing the sampling process.

Rendering technology is one of the core problems in the field of imagery. The problem to be solved is how to efficiently and high-quality calculate the propagation of light energy in the three-dimensional world to generate physically realistic pictures in a shorter time. However, the rendering of physical reality is very time-consuming and requires the use of Monte Carlo integration to solve the famous rendering equation \cite{kajiya1986rendering}, which usually requires a large number of sampling samples to achieve convergence. Various methods have been intensively studied to deal with this problem. In real-time rendering applications that require interactive frame rates, biased image spatial filtering (for displaying a single image) and light field spatial filtering (for displaying light fields) of noisy rendering results are common solutions. However, biased rendering is not suitable for design, physical simulation, training data generation for deep learning, and other high-quality rendering applications. In some rendering applications, time-domain discontinuities caused by mathematical biases, systematic errors, or small variances are unacceptable. At this time, the incident radiation field (light field) on the pixel can be reconstructed first, and mixed with the bidirectional reflection distribution function (BRDF) of the object surface as the probability distribution function (PDF) for guiding Monte Carlo sampling. If the reconstructed PDF is a good approximation of the true rendering function distribution, it is possible to generate mathematically unbiased rendering results by sampling a few Monte Carlo samples.

The academic community has been paying attention to light field adaptive sampling and reconstruction methods. Jensen et al. proposed a photon tracking algorithm to estimate the incident light field \cite{jensen1995importance}, and by calculating the histogram of the number of photons on the incident sphere, a probability function that can guide the path tracking sampling direction is generated. Based on this work, Budge et al. proposed a special PDF representation to efficiently render caustic effects \cite{budge2008caustic}. Vorba et al. proposed using an online training method to iteratively update the parameters of a Gaussian mixture model (GMM) to estimate the incident light field PDF\cite{vorba2014line}. Subsequent work improved this method based on the Russian roulette algorithm and extended it to the bidirectional scattering distribution function \cite{herholz2016product,vorba2016adjoint}. Hua et al. used a virtual point light source instead of a photon tracing algorithm to approximate the incident radiance\cite{hua2015guided}. Recently, Müller et al. proposed the use of a hybrid quadtree data structure in a spatially oriented four-dimensional space to represent online learnable incident light fields \cite{muller2017practical}. In addition, an online learning regression method is also used to sample the incident light field function for direct illumination to guide the sampling \cite{vevoda2018bayesian}.

\section{Monte Carlo Sampling and Reconstruction Using Reinforcement Learning}\label{sec:Monte}
The above-mentioned classic adaptive sampling and reconstruction method uses a prior-based manually designed model to perform online learning on a single scene to fit an unknown incident light field function. Recently, Huo et al. proposed to use cross-scene offline data to train deep neural networks \cite{huo2020adaptive} for better Monte Carlo adaptive sampling and reconstruction. This study proposes to train two different networks, called Quality Evaluation Network (Q-Network) and Reconstruction Network (R-Network), and use a large offline dataset to train both networks, where R-Net is a four-dimensional convolutional network, It is used to filter and reconstruct a sparsely sampled incident radiation field containing a lot of noise into a noise-free dense incident radiation field. The Q network is trained based on deep reinforcement learning strategies to guide the sampling process of obtaining sparse samples to improve sampling. s efficiency.

The entire drawing pipeline is described below. (a) First divide the image space into not small blocks. (b) The same orientation space shared by the pixels in each block, using a unified lattice structure to discretize the orientation space as initialization, and randomly generate sample points in each block. (c) Use the trained Q-network to predict the expected payoff of performing "sampling actions" within each block. The benefit refers to the quality improvement predicted by the network after sampling. When training the Q network, the quality improvement is obtained by comparing the output of the R network with the difference between the Ground Truth image. There are two kinds of "sampling actions". The first is re-sampling, which doubles the number of samples in the corresponding direction space of the block; the second is fine-sampling space, which divides the specified sampling space direction area into smaller granularity. grid to better reconstruct high frequency details within the region. After obtaining the predicted value map, the algorithm selects the action that can get the most benefit and executes it. (d) shows the new sampling orientation hierarchy obtained by applying the refinement sampling space to the initial orientation space. (e) Finally, the R network is used to reconstruct the sparse sampling points into a dense incident radiation field, which can be used for different applications such as path guidance or direct display.

The training method of the R network and the Q network is discussed below. The training dataset consists of 20 different scenes with randomly generated viewpoints and randomly modified scene lighting, materials and textures, resulting in a total of about 300 scenes. For each configuration, Ground Truth light field data with different sampling numbers and resolutions are generated, and the number of sampling points in each block is $(1, 2, 4, 8, 16) $ five kinds, including The resolution is $(16^1, 16^2, 16^3, 16^4, 16^5)$ five kinds. These data were first used to train the R network to converge and well reconstruct sparse samples into dense, noise-free light fields. The converged R network is used to train the Q network, which is also based on the same data set, but uses a stochastic reinforcement learning algorithm to simulate the data distribution of the actual sampling process to prevent data coupling. In order to make the Q network learn the benefits brought by a certain sampling action, the method sends the sampling point data before and after the sampling action into the R network at the same time to obtain two different reconstruction results. By comparing the reconstruction results with the Ground Truth data , it is possible to calculate the magnitude of the error reduction that the sampling action can bring, that is, the benefit. If an action yields higher returns, the trained Q-network guides the Monte Carlo sampling process to take that action first.

\section{Improvement}
The key limitation of the work to more generalized adaptive sampling and reconstruction applications is that it uses an explicit hierarchical tree to adapt for various resolutions. Given the recent advance in the implicit representation of radiance field, such as the famous NeRF \cite{mildenhall2020nerf}, it is possible to extend the light field representation to an implicit format. This modification yields several advancements. First, it reduces the limitation of adapting to the high-frequency light field due to the hierarchical data structure. Second, it might make the adaptive sampling more flexible and efficient since we do not need to infer each block separately. Finally, it might enable the reconstruction in higher-dimension space by recursively query of the implicit functions.




\section{Summarize}\label{sec:Conclusion}

In this manuscript, we discuss the possibility of using implicit radiance fields to overcome several difficulties in adaptive radiance field sampling. Other extended contents: 
\begin{enumerate}
    \item Deep Learning-Based Monte Carlo Noise Reduction By training a neural network denoiser through offline learning, it can filter noisy Monte Carlo rendering results into high-quality smooth output, greatly improving physics-based Availability of rendering techniques \cite{huo2021survey}, common research includes predicting a filtering kernel based on g-buffer \cite{bako2017kernel}, using GAN to generate more realistic filtering results \cite{xu2019adversarial}, and analyzing path space features Perform manifold contrastive learning to enhance the rendering effect of reflections \cite{cho2021weakly}, use weight sharing to quickly predict the rendering kernel to speed up reconstruction \cite{fan2021real}, filter and reconstruct high-dimensional incident radiation fields for unbiased reconstruction Drawing guide \cite{huo2020adaptive}, etc.
    \item The multi-light rendering framework is an important rendering framework outside the path tracing algorithm. Its basic idea is to simplify the simulation of the complete light path illumination transmission after multiple refraction and reflection to calculate the direct illumination from many virtual light sources, and provide a unified Mathematical framework to speed up this operation \cite{dachsbacher2014scalable}, including how to efficiently process virtual point lights and geometric data in external memory \cite{wang2013gpu}, how to efficiently integrate virtual point lights using sparse matrices and compressed sensing \cite{huo2015matrix}, and how to handle virtual line light data in translucent media \cite{huo2016adaptive}, use spherical Gaussian virtual point lights to approximate indirect reflections on glossy surfaces \cite{huo2020spherical}, and more.
    \item Automatic optimization of drawing pipelines Apply high-quality drawing technology to real-time drawing applications by optimizing drawing pipelines. The research contents include automatic optimization based on quality and speed \cite{wang2014automatic}, automatic optimization for energy saving \cite{ wang2016real,zhang2021powernet}, LOD optimization for terrain data \cite{li2021multi}, automatic optimization and fitting of pipeline drawing signals \cite{li2020automatic}, etc.
    \item Using physically-based process to guide the generation of data for single image refleciton removal \cite{kim2020single}; propagating local image features in a hypergraph for image retreival \cite{an2021hypergraph}; manging 3D assets in a blcok chain-based distributed system \cite{park2021meshchain}.
\end{enumerate}

{\small
\bibliographystyle{ieee}
\bibliography{Saliency}
}

\end{document}